\documentclass[12pt]{article}
\usepackage{graphicx} 
\usepackage[english]{babel}
\usepackage{amsmath}
\usepackage{anysize}
\usepackage{hyperref}
\usepackage{comment}
\usepackage{multirow}
\usepackage{appendix}
\usepackage[utf8]{inputenc}
\usepackage{t1enc}
\usepackage{setspace}
\usepackage{indentfirst}
\usepackage{amsmath, amsthm, amssymb}
\usepackage{epstopdf}
\usepackage{subfigure}
\usepackage{appendix}
\usepackage[bottom]{footmisc}
\usepackage{float}
\usepackage{cite}
\usepackage{underlin}
\usepackage{fancyhdr}
\usepackage{epigraph}
\usepackage{authblk}
\usepackage{url}
\marginsize{2cm}{2cm}{1.5cm}{1.5cm} 
\usepackage{tikz}

\title{Hydrodynamic description of direct photon spectrum and elliptic flow in Pb+Pb collisions at LHC}

\author{ Sándor Lökös$^{1,3}$, Gábor Kasza$^{2,3}$}
\affil{\small\textit{$^1$Institute of Nuclear Physics
Polish Academy of Sciences, Krak\'ow PL-31-342, Poland}}
\affil{\small\textit{$^2$HUN-REN Wigner Research Centre for Physics, Budapest H-1121, Hungary}}
\affil{\small\textit{$^3$Hungarian University of Agriculture and Life Sciences, Institute of Technology, Gyöngyös H-3200, Hungary}}
\date{\today}

\begin{document}

\maketitle

\begin{abstract}
In high energy heavy ion collisions a new state of matter, the strongly coupled quark gluon plasma is formed that exhibits similar properties as the Universe had couple of microseconds after the Big Bang, hence such collisions are usually referred as Little Bangs. Subsequent investigations showed that the created medium is a nearly perfect fluid whose time evolution can be described by hydrodynamic models. The distribution of the hadrons that are created in the freeze-out after a rapid expansion carry information about the final state. On the other hand, with penetrating probes, e.g., with direct photons, one can model the time evolution of the quark gluon plasma. In this paper, we present a hydrodynamic model that was inspired by an analytical solution of the equation of relativistic hydrodynamics. The invariant transverse momentum spectrum and the elliptic flow of direct photons were calculated, the model parameters were obtained by comparison the calculations to LHC ALICE data. Based on the the results we give an estimation for the initial temperature of the plasma.
\end{abstract}

\section{Introduction}

\label{sec:intro}

In the early 2000s, a series of RHIC experimental results showed that a new state of matter, the strongly coupled quark-gluon plasma (QGP) was created in heavy ion collisions~\cite{PHENIX:2001hpc,PHENIX:2000owy,STAR:2001eyo,PHENIX:2003qdw,PHENIX:2003qdj}. It was also determined that in this medium, the mean free path of the quarks is very small~\cite{PHENIX:2003qra} and its kinematic viscosity is close to the theorized minimum~\cite{PHENIX:2006iih,PHENIX:2010tme} hence relativistic perfect fluid models can be utilized to describe its time evolution as Landau has proposed in 1953~\cite{Landau:1953gs}. The reason for the wide applicability of hydrodynamics is that it has no internal physical scale: hydrodynamics can be applied from the femtometre distances of the particles to the largest scales of the Universe. 

Hadron spectra measured in heavy ion experiments carry information about the final state of the hot and dense hadronic medium but do not provide information about the time evolution of the quark-gluon plasma before the kinetic freeze-out~\cite{Csanad:2009sk}. We note that information about the phase after the chemical freeze-out but before the kinetic freeze-out might be accessible by measuring resonance ratios, see e.g. in Refs. \cite{Lokos:2022jze,PhysRevC.93.014911,BEBIE199295}. However, the length of this period is debated, but thermal models suggest that the chemical and the kinetic freeze-out of the hot and dense hadronic matter are two very close events in space-time, see for example Refs.~\cite{Broniowski:2001we,Broniowski:2002nf}.

Direct photon measurements can be considered as penetrating probes that can be utilized to gain information about the evolution of the medium. Direct photons are created continuously and they can reach the detectors without considerable  interaction with the plasma since they do not coupled strongly only electromagnetically and the cross section for such processes is small. Hence they are excellent probe to understand the time evolution of the QGP. Their spectrum is dominated by photons originated from hard scattering processes at high $p_{\rm T}$ and mostly thermal photons at low $p_{\rm T}$. Therefore the hydrodynamic calculations are considered to be valid for the lower $p_{\rm T}$ region, however, there is room for access of photons from other sources to this region too~\cite{PhysRevLett.109.202303,PhysRevC.90.014905,PhysRevC.105.014909,PhysRevC.92.054914}.

Direct photon spectra were measured by WA98 \cite{WA98:2000vxl}, PHENIX \cite{PhysRevLett.94.232301,PhysRevLett.104.132301,PhysRevLett.109.152302,PHENIX:2022rsx} and ALICE \cite{ALICE:2015xmh}. Direct photon flow were measured in PHENIX \cite{PhysRevLett.109.122302} and ALICE \cite{ALICE:2018dti}. Latter measurements attract great theoretical interest. The elliptic flow coefficient ($v_2$) of direct photons is of the same order of magnitude as for hadrons that is usually referred as \textit{direct photon puzzle}. Indeed, it is not well understood that how the strong photon emission at the early stages has as large $v_2$ as hadrons that are emitted at the later and cooler stage. There are sophisticated numerical models compared to data that underestimate the flow, however, can describe the spectrum \cite{PhysRevC.92.054914,PhysRevLett.114.072301,PhysRevC.96.014911}. 

In this manuscript, we calculate the direct photon spectrum and the $v_2$ elliptic flow coefficient based on a 1+3 dimensional boost-invariant perfect fluid hydrodynamic solution~\cite{Csorgo:2003ry}. We note that previously, using the above mentioned hydrodynamic solution of Ref.~\cite{Csorgo:2003ry}, direct photon spectra and the $v_2$ flow parameter have been calculated analytically in Ref.~\cite{Csanad:2011jq}. The calculation was performed using a second-order saddle-point approximation, and the resulting analytical formulae gave a statistically acceptable description of the direct photon spectrum measured by the PHENIX Collaboration~\cite{PHENIX:2011oxq}. More recently, an analytical formula for the thermal contribution of the direct photon spectrum was derived from a 1+1 dimensional hydrodynamic solution characterized by a locally accelerating velocity field~\cite{Kasza:2023rpx}. Comparing this analytical formula with the most recent spectrum of nonprompt direct photons measured by PHENIX~\cite{PHENIX:2022rsx}, the scaling behaviour of the direct photon spectrum was observed. However, this result is based on a 1+1 dimensional solution of perfect fluid hydrodynamics, hence the $v_2$ elliptic flow was not examined. We also discuss our model in comparison with kinetic approaches presented, e.g., in Refs.~\cite{Paquet:2015lta,Arnold:2001ms} and references in them.

The success of the analytical hydrodynamic models motivated us to compare our calculations with the ALICE Pb+Pb at $\sqrt{s_\textbf{\rm NN}}$=2.76 TeV data~\cite{ALICE:2015xmh,ALICE:2018dti}, with the aim of getting closer to solving the direct photon puzzle.


\section{The hydrodynamic model}
\label{sec:model}

A hydrodynamic model is defined by a source function that gives a parametrization of the spatio-temporal distribution of the created medium in the final state. It is taken to be the Bose-Einstein distribution on the four dimensional hypersurface that is parametrized by the Cooper-Frye flux term:
\begin{equation}
    S(x,p)d^4x = \mathcal{N} H(\tau) \frac{d\Sigma_{\mu} p^{\mu}}{\exp\left(\frac{p^{\mu}u_{\mu}}{T}\right)-1},
    \label{eq:source}
\end{equation}
where $p^{\mu}$ is the four-momentum of the photons, $u_{\mu}$ is the four-velocity of the fluid,  and $H(\tau)$ is a proper-time dependent window-, or opacity-function~\cite{Csorgo:1999sj} that we shall discuss later. The normalization factor $\mathcal{N}$ consists of the $g$ degeneracy factor and $\tau_{\rm R}$ characteristic time of the particle emission in the momentum space in a combination as $\mathcal{N}=\frac{g}{(2\pi \hbar)^3}\frac{1}{\tau_R}$. The $d\Sigma_{\mu} p^{\mu}$ term is called Cooper-Frye factor of the particle flux~\cite{Cooper:1974mv}, where $d\Sigma_{\mu}$ gives the normal vector of the hypersurface that is assumed to be parallel with the four-velocity, i.e., given in the form as
\begin{equation}
    d\Sigma^{\mu} = u^{\mu}dt\: d^3x.
\end{equation}

The $H(\tau)$ function describes the proper-time distribution of the opacity of the medium to photons. The photons are created continuously in the QGP and they do not participate in the strong interaction, thus their mean free path could be larger than the size of the thermalized fireball. Accordingly, $H(\tau)$ corresponds to the proper time duration of the thermal radiation, the time between the thermalization ($\tau_0$) and the freeze-out ($\tau_{\rm f}$) that can be described as
\begin{equation}
    H(\tau)=\Theta(\tau-\tau_{\rm 0})-\Theta(\tau-\tau_{\rm f}).
\end{equation}

The velocity field is taken to be a Hubble flow, i.e.,
\begin{equation}\label{eq:hubble}
u_{\mu}=\gamma\left(1,\textbf{v}\right)=\gamma\left(1,\frac{\textbf{r}}{t}\right),
\end{equation}
where \textbf{v} is the three-velocity, \textbf{r} is the three dimensional position vector and $t$ is the coordinate time. The Lorentz-factor is denoted by $\gamma$.

In this paper, we describe the invariant transverse momentum distribution and the elliptic flow of the direct photons therefore we give the model in cylindrical spatial coordinates that suit better the geometry of the system than the Cartesian coordinates. The two sets of coordinates are related as $r=\sqrt{r_x^2 + r_y^2},\: \alpha = \arctan\left(r_y/r_x\right)$. Hence the position and momentum four-vectors have the form as
\begin{equation}
x_\mu =(t,r,\alpha,r_z),\hspace{0.3cm} p_\mu = (E,p_{\rm T}\cos(\phi),p_{\rm T}\sin(\phi),p_z)
\end{equation}
where $r$ is the radial coordinate, $\alpha$ is the spatial azimuth angle, $r_z$ is the coordinate in the beam direction, $E$ is the photon's energy, $p_{\rm T}$ is the transverse momentum, $\phi$ is the momentum azimuth angle and $p_z$ is the longitudinal component of the momentum. Taking into account for the limited pseudorapidity acceptance of the ALICE detector ($|\eta|<0.8$) further simplification can be made as $p_z \approx 0$. Therefore, in cylindrical coordinates,
\begin{equation}
    p^{\mu}u_{\mu} \approx p_{\rm T} \gamma \left[1 - r/t\cos(\phi-\alpha)\right].
\end{equation}
As it can be seen from Eq.~(\ref{eq:source}), the temperature profile still needs to be given. We shall discuss that in Sec.~\ref{sec:tempprof}. Regardless of the temperature profile the observable quantities like the invariant momentum distributions or the elliptic flow can be obtained by calculating the space-time moments of the source function. The invariant momentum distribution is given as (taken at mid-rapidity: $y=0$)
\begin{equation}
    N_1(p_{\rm T},\phi) = E\left.\frac{d^3 N}{dp^3}\right|_{p_z=0}=\left.\frac{d^3 N}{d\phi dp_{\rm T} dy}\right|_{y=0} = \int S d^4 x.
\end{equation}

One of the observable quantity we calculate is the transverse momentum distribution at mid-rapidity. It can be calculated from the invariant momentum distribution by integrating $N_1(p_{\rm T},\phi)$ with respect to the azimuth angle:
\begin{equation}
    \left.\frac{d^2N}{p_{\rm T} dp_{\rm T} dy}\right|_{y=0} = \int\limits_{0}^{2\pi}d\phi N_1(p_{\rm T},\phi).
\label{eq:N1}
\end{equation}
The anisotropies of the invariant momentum distribution function are defined as $n$-th order Fourier coefficients with respect to the angle of the $n$-th order event plane. In this paper, we restrict ourselves to investigate the second order anisotropies:
\begin{equation}
    v_2(p_{\rm T}) = \frac{\int\limits_{0}^{2\pi}d\phi 
    \cos(2\phi) N_1(p_{\rm T},\phi)}{\int\limits_{0}^{2\pi}d\phi N_1(p_{\rm T},\phi)}.
\label{eq:v2}
\end{equation}

It could be emphasized that our model differs from approaches based on relativistic kinetic theories (e.g. Refs.~\cite{Paquet:2015lta,Arnold:2001ms}). Instead of calculating the photon spectrum from the rates of the photon-emitting interactions in the thermalized medium we employ a collective source function for the photons with the assumption of Bose-Einstein distribution as its shape that is determined by the dynamics and temperature of the thermalized medium. This approach results in a semi-analytical model for both the spectrum and the elliptic flow. Nonetheless, we estimated the difference between the two approaches by adopting a similar technique that was detailed in Ref.~\cite{Csanad:2011jq} as follows. According to several microscopic models, the rate of a process of type $A+B\rightarrow X$ can be written as $n_A n_B\langle v \sigma\rangle$, where $n_A$ and $n_B$ are the particle densities of particle types $A$ and $B$. The cross-section of the process is denoted by $\sigma$ and $v$ stands for the incoming velocity. In our calculations, we employed the hydrodynamic solution given in Ref.~\cite{Csorgo:2003ry}, where $n\propto T^3$. Utilizing it in the equation above, we obtain $n_A n_B\propto T^6$, hence the photon production rate is proportional to $T^6$. In our model, the integral of the source function with respect to the momentum is proportional to $T^2$, thus the production rate of thermal photons will be proportional to $T^4$ instead of $T^6$. Accordingly, by multiplying the source function by $T/T_0$, we can estimate the difference between the two approaches, but since the incoming velocity $v$ could also depend on the temperature, we repeated our analysis with prefactor $(T/T_0)^2$, too, and taken into account for their effect as a systematic model choice uncertainty. The magnitude of the uncertainty of the initial temperature stemming from this choice is $<2\%$. It is also included in the systematic uncertainties of the fit parameters in Tab.~\ref{tab:ALICE_fit_results}.

\section{Temperature profile}
\label{sec:tempprof}

In the source function defined in Eq.~(\ref{eq:source}), the temperature has to be specified. In this work, the temperature is given by an exact and analytic, self-similar solution of perfect fluid hydrodynamics based on Ref.~\cite{Csorgo:2003ry}. Although this solution was found with finite chemical potential, in this manuscript we describe the medium at zero chemical potential. In such a medium, defining conserved particle number is problematic due to the continuously created and annihilated virtual particles, so the solution used in this manuscript can be given by the temperature and the entropy density with the equation of state $\varepsilon=\kappa p$, where $\kappa$, in general, can be the function of the temperature. 

The spatial anisotropy can be introduced into this hydrodynamic model by the so-called scale variable $s$, that has a vanishing co-moving derivative $u^{\mu}\partial_{\mu}s=0$. In the case of an elliptic fireball, it can be written in elliptical coordinates as
\begin{equation}
    s = \frac{r^2}{R^2} \left( 1+\epsilon_2\cos(2\alpha) \right) + \frac{r_z^2}{Z^2},
\end{equation}
where the parameters $R$ and $\epsilon_2$ are related to the axis of the ellipsoid in the Cartesian system as
\begin{equation}
    \frac{1}{R^2}=\left( \frac{1}{X^2} + \frac{1}{Y^2} \right), \hspace{1cm} \epsilon_2 = \frac{Y-X}{Y+X}.
\end{equation}
The scales are denoted by $X, Y$ and $Z$, and they stand for the time dependent axes of the fireball. Here we use $X(t)=\dot{X}_0 t$, $Y(t)=\dot{Y}_0 t$ and $Z(t)=\dot{Z}_0 t$ as described in the boost-invariant solution of Ref.~\cite{Csorgo:2003ry}. The $s$ dependence of the temperature is carried by an arbitrary function $\mathcal{T}(s)$, so in our model the temperature is factorized into the function $\mathcal{T}(s)$ and a homogeneous term that depends only on the proper-time as $T(\tau,s)=T_{\rm H}(\tau)\mathcal{T}(s)$. This is due to the definition of $s$ which is given by the $u^{\mu}\partial_{\mu}s=0$ equation, thus the $s$ dependence in the temperature is eliminated from the equation of energy conservation~\cite{Csorgo:2003ry}. If the velocity field of the fluid is defined by Eq.~(\ref{eq:hubble}) and the EoS is written as $\varepsilon=\kappa p$, the solution of the temperature can be calculated from the following differential equation, which corresponds to the energy conservation:
\begin{equation}\label{eq:temp_eq}
    \left[\left(1+\kappa\right)\frac{d}{dT}\left(\frac{\kappa T}{1+\kappa}\right)\right]\frac{\partial T}{\partial \tau} + \frac{3T}{\tau} = 0.
\end{equation}
This differential equation can be solved in three cases:
\begin{enumerate}
    \item The temperature depends only on $\tau$ and $\kappa$ can be temperature dependent: $T\equiv T(\tau)$, $\mathcal{T}(s)=1$, $\kappa\equiv\kappa(T)$.
    \item The coefficient of the derivative of $T$ with respect to $\tau$ in Eq.(\ref{eq:temp_eq}) is fixed to a constant, so the temperature can depend both on $\tau$ and $s$, and $\kappa$ is a temperature dependent function which is given in Ref.~\cite{Csorgo:2016ypf}.
    \item The temperature can depend on both $\tau$ and $s$, but $\kappa$ has to be independent from the temperature: $T(\tau,s)$, $\kappa=\textnormal{constant}$.
\end{enumerate}

The first option has the benefit that it allows us to incorporate the lattice QCD EoS (e.g. that was published in Ref.~\cite{Borsanyi:2010cj}) into our calculations. However, in this case, Eq.(\ref{eq:temp_eq}) allows only the $\tau$ dependence of the temperature, so there is no spatial decay of the temperature of the fireball. This case clearly contradicts the intuitive picture of the expanding and cooling fireball from heavy ion collisions and, indeed, does not give a satisfactory description of the measured spectra. However, we note that an implicit hydrodynamic solution considering spherical source was derived using a class of EoS that includes the lattice QCD EoS~\cite{Csanad:2012hr}.

The second option allows one to introduce spatially decreasing profile for the temperature using the arbitrary function $\mathcal{T}(s)$, and in this case too, the EoS belongs to the same class as the EoS of the lattice QCD. However, the form of $\kappa(T)$ is then fixed and only applicable over a certain range of temperature, as discussed in Ref.~\cite{Csorgo:2016ypf}.

In the third option, which is relevant in this work, $\kappa$ can only be a temperature-independent constant, which can be considered as a value averaged over the evolution of the fireball. In the nonrelativistic limit of the solution discussed here, we observed that the application of the temperature dependent EoS has a negligible effect in the fireball expansion dynamics, so for simplicity, $\kappa$ is considered to be constant and we utilize the solution for the temperature from Ref.~\cite{Csorgo:2003ry} in the form of
\begin{equation}
    T(\tau,s)=T_{\rm f} \left(\frac{\tau_{\rm f}}{\tau}\right)^{3/\kappa}\mathcal{T}(s).
    \label{eq:Ttau}
\end{equation}
Here $T_{\rm f}$ is the freeze-out temperature and $\tau_{\rm f}$ is the freeze-out proper time. Since we consider vanishing bariochemical potential, the constant proportionality factor $\kappa$ between the pressure and energy density is equal to the inverse square of the speed of sound: $\kappa=c_s^{-2}$. The scaling function, $\mathcal{T}(s)$, is arbitrary and here taken to has a Gaussian profile:
\begin{equation}
    \mathcal{T}(s)=\exp(-bs/2),
\end{equation}
where we introduced a new model parameter, $b$, which is an arbitrary positive real number. It is worth noting that the value of $b$, $\dot{R}_0$ and $\dot{Z}_0$ cannot be determined unambiguously when comparing our model with the data, since the temperature profile and hence our integral formula developed to describe the direct photon spectrum depend only on the ratios of $b/\dot{Z}_0$, and $b/\dot{R}_0$.

\section{Model comparison to ALICE data}

Observables that can be calculated from the model and compared to data are described above in Eq.~(\ref{eq:N1}) and Eq.~(\ref{eq:v2}). Because of the rather complicated dependence on the coordinates, numerical integration is employed. However, in the same model we use, Eq.~(\ref{eq:N1}) and Eq.~(\ref{eq:v2}) have been analytically integrated earlier in Refs.~\cite{Csanad:2011jq,Csanad:2011jr}, but only at the cost of several approximations.

Direct photon measurements are experimentally challenging therefore limited amount of data is available. We choose to compare our model calculations to single particle transverse momentum \cite{ALICE:2015xmh,hepdata.73093} and elliptic flow \cite{ALICE:2018dti,hepdata.88050} data simultaneously that were measured by ALICE at $\sqrt{s_\textmd{NN}}=2.76$ TeV in Pb+Pb collisions in the 0-20\% centrality class.

We compare our calculations to the data by standard $\chi^2$ minimization method employing ROOT Minuit2 package~\cite{ROOT}. Only the statistical uncertainties of the data were considered in the minimization process. To reach sufficient precision in the numerical integration, considerable amount of computational time was necessary therefore we restrict our fit to five parameters, namely, to the freeze-out temperature $T_{\rm f}$, freeze-out proper-time $\tau_{\rm f}$, ellipticity $\epsilon_2$, the EoS $\kappa$ and normalization $\mathcal{N}$ as it is listed in Tab. \ref{tab:ALICE_fit_results}. The remaining parameters were fixed to values that are compatible with available results in the literature, e.g., in Ref. \cite{Csanad:2011jq}. The statistical uncertainties of the fit parameters were determined by the Minuit2 package.

We also determined the systematic uncertainties of our choice of the fixed parameter values. The systematic checks were done as follows: we redo the fits by changing the value of one fixed parameter at a time to a larger or smaller value taking $\sim 10\%$ variation and obtain new values for the fit parameters as alternative ones.

Then we calculate the relative deviation of the alternative values from the default one. Repeating this for the fixed parameters, the total systematic uncertainties of the fit parameters were calculated as the quadratic sum of the relative deviations. Since the parameters could be correlated which would reduces their contribution to the systematic uncertainties, the utilized method can be regarded as a conservative estimate.

\begin{figure}[h!]
\centering
\resizebox{0.95\textwidth}{!}{%
  \includegraphics{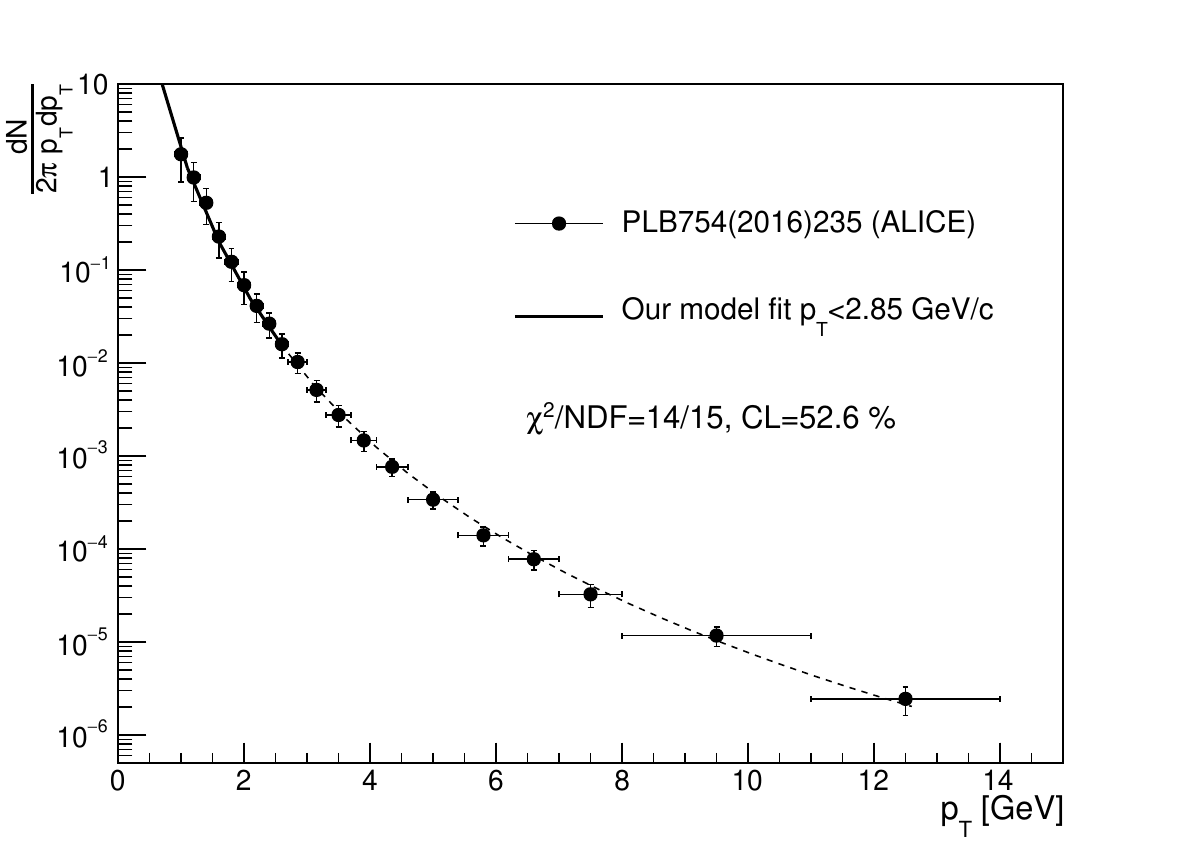}
}
\caption{Model fit to the direct photon spectrum measured by the ALICE Collaboration \cite{ALICE:2015xmh,hepdata.73093}  The fit was made simultaneously for the spectrum and the elliptic flow for point with $p_T < 2.85$ GeV/c shown with solid line but we show the model calculation up to $p_T<14$ GeV/c with dashed line.}
\label{fig:ALICE_direct_photon_spectrum_fit}
\end{figure}

\begin{figure}[h!]
\centering
\resizebox{0.95\textwidth}{!}{%
  \includegraphics{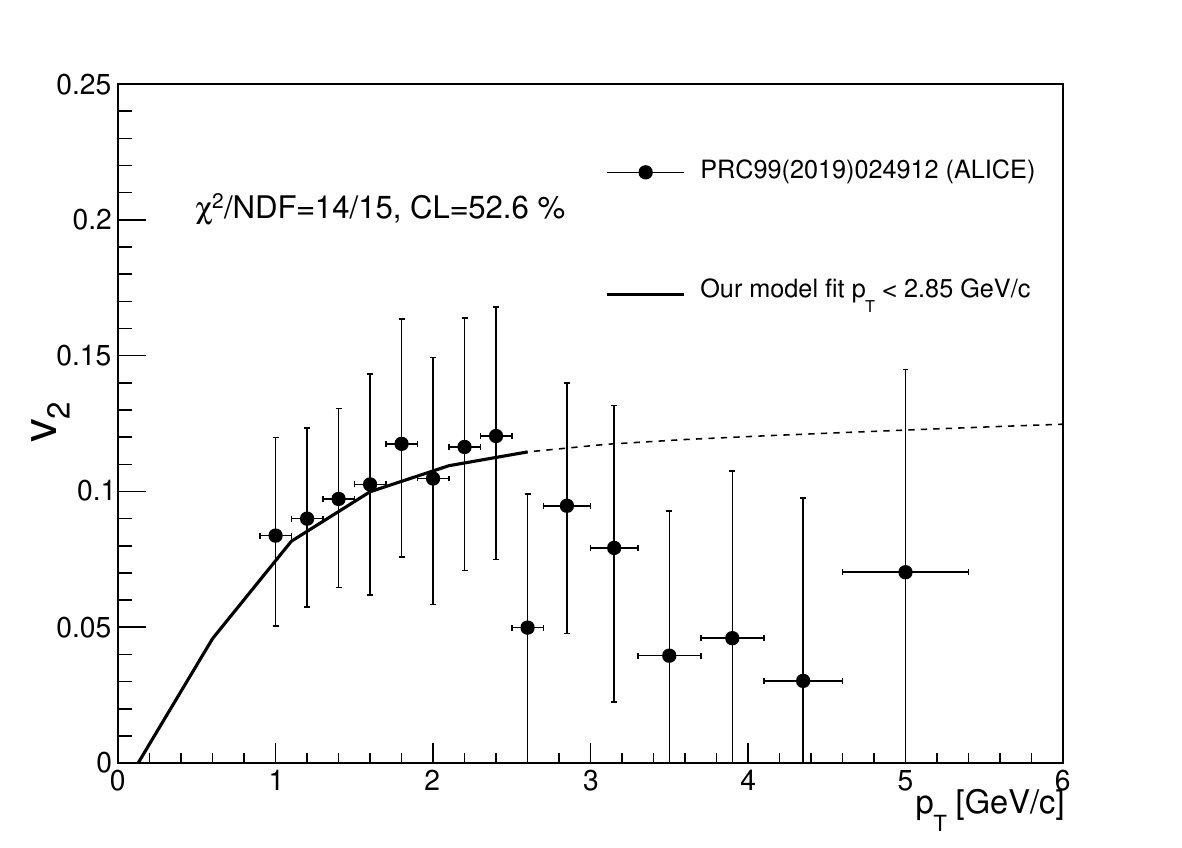}
}
\caption{Model fit of the direct photon elliptic flow measured by the ALICE Collaboration \cite{ALICE:2018dti,hepdata.88050}. The fit was made simultaneously for the spectrum and the elliptic flow for point with $p_T < 2.85$ GeV/c shown with solid line but we show the model calculation up to $p_T=6$ GeV/c with dashed line.}
\label{fig:ALICE_direct_photon_v2_fit}
\end{figure}

Our fit results are shown in Fig. \ref{fig:ALICE_direct_photon_spectrum_fit} for the spectrum and in Fig.~\ref{fig:ALICE_direct_photon_v2_fit} for the elliptic flow. We also summarize the fit results in Tab.~\ref{tab:ALICE_fit_results}. 


\section{Discussion}

Our simple hydrodynamic model gives a statistically acceptable simultaneous description of the $p_{\rm T}$-spectrum and the elliptic flow of direct photons in ALICE Pb+Pb at $\sqrt{s_\textmd{NN}}=$ 2.76 TeV collisions with 0-20\% centrality in terms of $\chi^2/$NDF, even beyond the expected limit of validity of the hydro description that is usually considered to be $p_{\rm T}<$ 2-3 GeV. Our calculations show a good agreement with the data, however the large uncertainties of the experimental data allow to draw conclusions only on the qualitative level.
    
Based on the temperature profile that we described in Section~\ref{sec:tempprof}, we were able to determine the initial temperature of the plasma in the center of the fireball, that is 
\begin{equation}
T_\textmd{init}{=}T(\tau{=}1,s{=}0){=}418{\pm}31(\textmd{stat}){\pm}35(\textmd{syst}) \textnormal{ MeV}.
\label{eq:Tinitnew}
\end{equation}
This result for the initial temperature is clearly higher than the Hagedorn temperature, even if we take into account that theoretical predictions give a wide range for the Hagedorn temperature~\cite{Broniowski:2000bj,Broniowski:2004yh,Cohen:2011cr,Cleymans:2011fx}. Hence we conclude that the medium created in Pb+Pb at $\sqrt{s_{NN}}=$2.76 TeV collisions cannot be interpreted by a description involving hadrons only. In addition, we note that our results provide values for the initial and freeze-out temperatures (or equivalently for the initial and freeze-out proper times) that are compatible with the results of other analyses on hadronic~\cite{Csanad:2009wc,Ze-Fang:2017ppe} and direct photon~\cite{PHENIX:2008uif,Nayak:2008tk,Nayak:2012dj,Csanad:2011jr,Csanad:2011jq,Kasza:2023rpx} observables.

\section{Summary}

In our studies, we investigate the applicability of the hydrodynamic description of the direct photon data measured by ALICE experiments at LHC. In our model, we utilized an equation of state characterized by the average value of the speed of sound with an exact and analytical solution of relativistic hydrodynamics. We found a good agreement between the data and the model with realistic parameters. We were able to determine the initial temperature of the fireball. Our model calculations are closer to the data than those hydrodynamics and transport approaches that were compared to the ALICE data in Ref.~\cite{ALICE:2018dti}.

\section{Acknowledgement}
This research was funded by the NKFIH grants K-138136 and K-147557, and the KKP-2024 Research Excellence Programme of MATE, Hungary. The authors would like to express their gratitude to prof. Giorgio Torrieri, prof. G{\'a}bor D{\'a}vid and prof. Tam{\'a}s Cs{\"o}rg{\H o} for their insightful comments.

\newpage

\begin{table}[h!]
\centering
\begin{tabular}{cccc}
\hline\noalign{\smallskip}
Parameter &  Fit value & Stat. uncert. & Syst. uncert.\\
\noalign{\smallskip}\hline\noalign{\smallskip}
$T_{\rm f}$ [MeV]       &  121.2  &  0.02   &  5.2  \\
$\tau_{\rm f}$ [fm]     &  5.22   &  0.52   &  0.40  \\
$\epsilon_2$            &  0.29   &  0.01   &  0.08  \\
$\kappa$                &  4.01   &  0.09   &  0.01  \\
$\mathcal{N}$           &  0.07   &  0.02   &  0.01  \\
$b/\dot{R}_0$           &  0.53    &  const. & const. \\
$b/\dot{Z}_0$           &  0.83    &  const. & const. \\
\noalign{\smallskip}\hline
\end{tabular}
\caption{Obtained parameter values from the simultaneous fit to the ALICE data of direct photon spectrum \cite{ALICE:2015xmh} and direct photon elliptic flow \cite{ALICE:2018dti}. The fixed parameters are indicated with const. in the uncertainties columns. The systematical uncertainties are determined based on fits with fixed parameters varied by $\sim \pm 10\%$ around their nominal value.}
\label{tab:ALICE_fit_results}
\vspace*{5cm}
\end{table}

\end{document}